\documentclass[twocolumn,
aps,
prb,
amsmath,
amssymb,
floatfix,
noeprint,
reprint]{revtex4-2}

\usepackage{graphicx,xcolor,mathtools,bm,MnSymbol}
 \usepackage{soul}    
 \usepackage{nicematrix}
\usepackage{xcolor}
\usepackage{physics}
\usepackage{verbatim}
\usepackage{verbatim}
\usepackage{esint}
\usepackage{enumitem}
\usepackage{bbold}

\def \beq {\begin{eqnarray}}
\def \eeq {\end{eqnarray}}

\definecolor{DarkRed}{RGB}{100,0,0}

\definecolor{DarkGreen}{RGB}{0,100,0}

\usepackage[bookmarks=false,linkcolor=blue,urlcolor=blue,colorlinks,citecolor=blue]{hyperref}
\usepackage[all]{hypcap}

\usepackage[english]{babel}
\usepackage[babel,kerning=true,spacing=true]{microtype}

\begin{document}

\title{Robust Fermi liquid instabilities in sign problem-free models}

\author{Ori Grossman}
\email{ori.grossman@weizmann.ac.il}
\author{Erez Berg}
\affiliation{Department of Condensed Matter Physics, Weizmann Institute of Science, Rehovot, 76100, Israel}

\begin{abstract}
Determinant Quantum Monte Carlo (DQMC) is a powerful numerical technique to study many-body fermionic systems. 
In recent years, several classes of 
sign-free (SF) models have been discovered, where the notorious sign problem can be circumvented. However, it is not clear what are the inherent physical characteristics and limitations of SF models. In particular, which zero-temperature quantum phases of matter are accessible within such models, and which are fundamentally inaccessible? 
Here, we show that a model belonging to any of the known SF classes within DQMC cannot have a stable Fermi liquid ground state in spatial dimension $d\ge 2$, unless the anti-unitary symmetry that prevents the sign problem is spontaneously broken (for which there are currently no known examples in SF models).
For SF models belonging to one of the symmetry classes (where the absence of the sign problem follows from a combination of non-unitary symmetries of the fermionic action), any putative Fermi liquid fixed point generically includes an attractive Cooper-like interaction that destabilizes it.  
In the recently discovered lower-symmetry classes of SF models, the Fermi surface is generically unstable even at the level of the quadratic action. 
Our results suggest a fundamental link between Fermi liquids and the fermion sign problem. Interestingly, our results do not rule out a non-Fermi liquid ground state with a Fermi surface in a sign-free model.
 \end{abstract}

\maketitle

{\it Introduction.---}
The importance of reliable and practical simulations of strongly correlated fermionic systems cannot be overstated. 
In recent decades, substantial progress was made 
thanks to the development of the Determinant Quantum Monte Carlo (DQMC) technique~\cite{Scalapino1981,blankenbecler1981sugar,Hirsch1981,hirsch1983discrete,White1989,assaad2008}.
However, this technique is often hindered by the sign problem~\cite{Loh1990,pan2022sign}, 
associated with negative or complex amplitudes in the quantum partition sum. The sign problem results in inefficient simulations, generally scaling exponentially with system size and inverse temperature~\cite{Troyer2005,assaad2008} (although interesting exceptions exist~\cite{Zhang2022,Hong2022}).  

Interestingly, certain classes of fermionic models do not suffer from the sign problem~\cite{wu2005,Li2015,Wang2015,Wei2016,Wei2017,Li2016,Li2019}, and hence can be solved at polynomial cost. In these models, the absence of the sign problem is guaranteed by a combination of symmetries of the fermionic action. Sign free (SF) models were used to simulate a plethora of interesting phenomena, including fermionic quantum criticality~\cite{Assaad2013,Gerlach2017,Berg2019,Grossman2021,Liu2022,Schwab2022,Lunts2022} and unconventional superconductivity~\cite{Assaad1996,Schattner2016a,Li2017,Xu2021}.  
These developments raise the question of the physical properties and intrinsic limitations of sign problem free models~\cite{Ringel2017,Golan2020,Smith2020,Hong2022,Mondaini2022}.  
In particular, which quantum phases of matter can be accessed within SF models, and which are fundamentally inaccessible?

In this work, we show that all the currently known classes of sign problem free models within DQMC cannot support a stable Fermi liquid (FL) ground state. 
This is shown by demonstrating that any putative Fermi surface in a SF model is generically unstable in the presence of interactions at zero temperature. Depending on the model, the instability may be towards a fully gapped superconducting or density wave state, or towards a Dirac semi-metal~\cite{Pujari2016,Gazit2017}.

It is important to note that we cannot rule out a fine-tuned SF model with a Fermi surface (e.g., a model of non-interacting electrons). Our argument shows that in such cases, the Fermi surface is unstable, in the sense that a generic, arbitrarily small perturbation can destroy it~\footnote{This is in contrast, e.g., to a superfluid phase, which is realizable within a SF model and is robust against arbitrary small perturbations (both ones that preserve the SF property and ones that do not).}.
In addition, a non-FL state with a Fermi surface (of the type that arises, e.g., at certain quantum critical points) is not ruled out.

\begin{table}
\centering
   \caption{The known sign-free (SF) classes (see text for a detailed description), and the origin of the Fermi surface instability in each class. In the symmetry classes (Symm. SF), there is an interaction-induced instability of the Fermi surface that opens a gap through spontaneous symmetry breaking. 
   In the lower-symmetry classes (lower-symm. SF), the Fermi surface is not protected (in general) by symmetry, and a gap generically opens even at the single-particle level.}
\includegraphics[width=1\columnwidth]{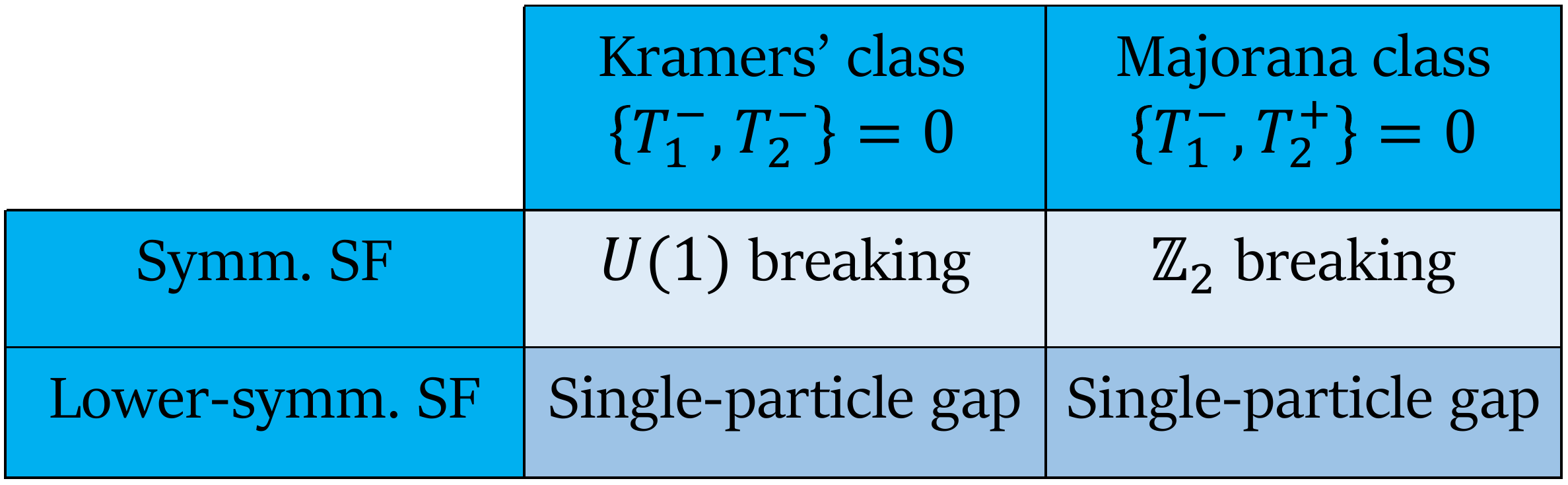}
    \label{table:classes} 
\end{table}

Our main findings are summarized in Table~ \ref{table:classes}. In the symmetric SF classes~\cite{Li2019} the SF property is guaranteed by the combination of two anti-unitary time-reversal symmetries (TRSs) of the fermionic action matrix. A detailed description of the different SF classes is given below. The product of the two TRSs defines a unitary symmetry, which is either a $U(1)$ or a $\mathbb{Z}_2$ symmetry, depending on the class. Interactions generically lead to spontaneous breaking of the unitary symmetry, gapping out the Fermi surface. There are many known examples of these phenomena (see Suppl. Material \footnote{See the appendices below, where we  provide detailed examples and explain proof steps that require further clarification.}); our work shows that these are general properties of SF models. There are also lower-symmetry classes where the SF is due to  less strict conditions~\cite{Wei2016,Wei2017}. We show that in these cases, generically, there is no FS even at the single-particle level, since there is no symmetry to protect it. If a symmetry that protects a FS is imposed, the FS is unstable in the presence of interactions, as in the symmetric classes. 

{\it Sign-free DQMC.---}
 DQMC is based on introducing  a bosonic field $\phi$ via a Hubbard-Stratonovich (HS) transformation, deriving an effective action for $\phi$ by integrating out the fermions, and averaging 
 stochastically over $\phi$. The field $\phi$ mediates the fermionic interaction~\cite{assaad2008}. Alternatively, $\phi$ may represent a physical boson (such as a phonon). A typical action 
 has the form of  $S=S_{F}+S_{\phi}+S_{\rm{Int}}$ 
where  $S_F$ is the non-interacting fermionic action, $S_{\phi}$ is the bosonic action, and $S_{\rm{Int}}$ is a Yukawa-like interaction. We assume that $S_\phi\in\mathbb{R}$. Upon integrating out the fermions, an effective bosonic action is obtained: $S'_{\phi} = S_{\phi}+\ln{\det M_{\phi}}$, where $M_{\phi}$ is the ($\phi$ dependent) quadratic fermionic action matrix, $\bar{\psi}M_{\phi}\psi =S_{\rm{Int}}+S_{F}$.  
  The fermionic problem has therefore been mapped to a classical statistical mechanical problem in $d+1$ dimensions for the field $\phi(\mathbf{r},\tau)$.
  
However, it is not guaranteed that the statistical weights in the partition sum can be treated as probabilities, since they are not necessarily  real and non-negative. 
This is known as the sign problem. 
Models that satisfy $S'_{\phi}\in \mathbb{R}$ (or equivalently $\det M_{\phi} \geq 0$) for any configuration $\phi(\mathbf{r},\tau)$ are known as SF models.

A set of sufficient conditions is known to guarantee the absence of the sign problem~\cite{Li2019}. These conditions are most conveniently stated using a Majorana representation, writing the complex fermion field in terms of two real (Majorana) fields: $\psi=\frac{1}{2}(\gamma_1 + i\gamma_2)$. The fermionic bilinear action takes the form $\gamma^{T}\tilde{M}_{\phi}\gamma$ where $\gamma=(\gamma_{1,1},...\gamma_{1,N},...\gamma_{2,N})$ (for $N$ Dirac fermions) and $\tilde{M}_{\phi}$ is a $2N\times 2N$ skew-symmetric matrix. In addition, a Majorana TRS, $T$, is defined as an anti-unitary operator that satisfies $\left[\tilde{M}_{\phi},T\right]=0$ for any $\phi(\mathbf{r,\tau})$. In this framework, one can distinguish between two fundamental SF classes: 

\begin{itemize}
    \item \textbf{Kramers' class:} $\tilde{M}_{\phi}$ has two mutually anti-commuting TRSs, satisfying $(T_1^{-})^2=(T_2^{-})^2=-1$. Since $(iT_1T_2)^2=1$, models of this class have a conserved $U(1)$ charge, $\hat{Q}=\gamma^{T} iT_1 T_2\gamma$. They can therefore be represented by Dirac fermions with a $U(1)$ symmetry ($\gamma^{T}\tilde{M}_{\phi}\gamma \to \bar{\psi}M_{\phi}\psi$). By Kramer's theorem, the eigenvalues of $M_\phi$ come in complex conjugate pairs.
     \item \textbf{Majorana class:} $\tilde{M}_{\phi}$ has two mutually anti-commuting TRSs, satisfying $(T_1^{-})^2=-1,(T_2^{+})^2=1$. 
  Since $U=T_1^{-}T_2^{+}$ is a $\mathbb{Z}_2$ unitary symmetry, $\tilde{M}_{\phi}$ can be brought to the block diagonal form 
  \begin{align}
      \tilde{M}_{\phi}= \begin{bmatrix}
        B & 0 \\ 0 & B^{*} 
      \end{bmatrix}
      \label{Eq:Majorana_matrix1}
  \end{align}
 in the eigenbasis of $U$. Integrating out the fermions yields $\mathrm{Pf}(\tilde{M}_\phi) = \mathrm{Pf}(B)\mathrm{Pf}(B^*)\ge 0$ (where $\mathrm{Pf}(B)$ is the Pfaffian of the skew-symmetric matrix $B$). 
  Models of this type may not have a $U(1)$ symmetry, but can still have a FS of the zero energy Bogoliubov-like excitations, protected at the single-particle level by the $\mathbb{Z}_2$ symmetry $U$. 
\end{itemize}

The conditions above can be somewhat relaxed~\cite{Wei2016,Wei2017}. In the so called lower-symmetry classes, 
we keep the requirement
 $\left[\tilde{M}_{\phi},T_{2}^{\pm}\right]=0$, but the second condition becomes  
$iK\left[ T_{1}^{-},\tilde{M}_{\phi}\right]  \leq 0$  (i.e, the left-hand side is a negative semi-definite matrix), where $K$ is complex conjugation. As before $\left\{T_{1},T_{2}\right\}=0$. 
 These requirements are sufficient to guarantee $\det M_{\phi}\geq 0$ ~\cite{Wei2017}.  
 The limiting case where $\left[\tilde{M}_{\phi},T_{1}^{-}\right]=0$ corresponds to the symmetry classes discussed above. In case of a strict inequality, 
 we have two lower-symmetry SF classes: the Majorana class ($T_{2}^{+}$) and the Kramers' class ($T_{2}^{-}$).

 {\it Fermi surface instability.---}
We consider a general, translationally invariant model in $d\geq 2$ spatial dimension, that belongs to one of the SF classes. 
The model includes Majorana fermions interacting with a bosonic field via a Yukawa coupling (note that any model of fermions with a quartic interaction can be recast in this form via a HS transformation). 
The lattice-scale, ultra-violet (UV) action is given by
\begin{align}
   S_{\rm{UV}}=\int \mathrm{d}{k}\  \overline{\gamma}_{\alpha,{k}}
  \frac{\left(i\omega\delta_{\alpha,\alpha'} -h_{\alpha,\alpha',\mathbf{k}}  \right)}{2} \gamma_{\alpha',{k}}
+S_{\phi}+S_{\rm{Int}},
\label{Eq: UV action}
\end{align} 
where $\alpha$ denotes a general fermionic flavor such as spin or Majorana flavor ($\gamma_{1,2}$), and summation over repeated indices is assumed. We have defined $\overline\gamma_{k}\equiv \gamma_{-k}$, and used the notation $k=\{i\omega,\mathbf{k}\}$, where $\mathbf{k}$ is the spatial momentum.   
$S_{\phi}$ is a bosonic action, and $S_{\rm{Int}}$ is given by
\begin{equation}
    S_{\rm{Int}}=\int \mathrm{d} k \mathrm{d} k' \frac{1}{2}\lambda^{\nu}_{\alpha,\alpha',{k},{k'}}
\phi^{\nu}_{{k},{k'}}  \overline{\gamma}_{\alpha,{k}}  \gamma_{\alpha,{k}'} .
\end{equation}
Here $\nu$ is the index of the auxiliary field  and $\lambda$ is the coupling function (which depends on  momenta, bosonic index, and fermionic flavor). 
Note that the bosonic field $\phi^{\nu}$ may also complex, but $S_{\phi}\in\mathbb{R}$ as the model is SF. In addition, we assume that our action corresponds to a physical Hamiltonian. I.e., Eq.(\ref{Eq: UV action}) corresponds to a hermitian Hamiltonian $\hat{H}(\hat{\gamma},\hat{\phi},\hat{\pi})$, where $\hat{\phi},\hat{\pi}$ are canonically conjugate operators.

We use proof by contradiction, assuming that the ground state is a FL, and showing that the putative FL phase is unstable. 
The proof proceeds in two steps: (1) Obtaining the low-energy FL effective action, 
(2) Showing that the SF requirement necessitates the existence of an instability of the Fermi surface. 

\begin{figure}
        	\centering\includegraphics[ width=0.8\columnwidth ]{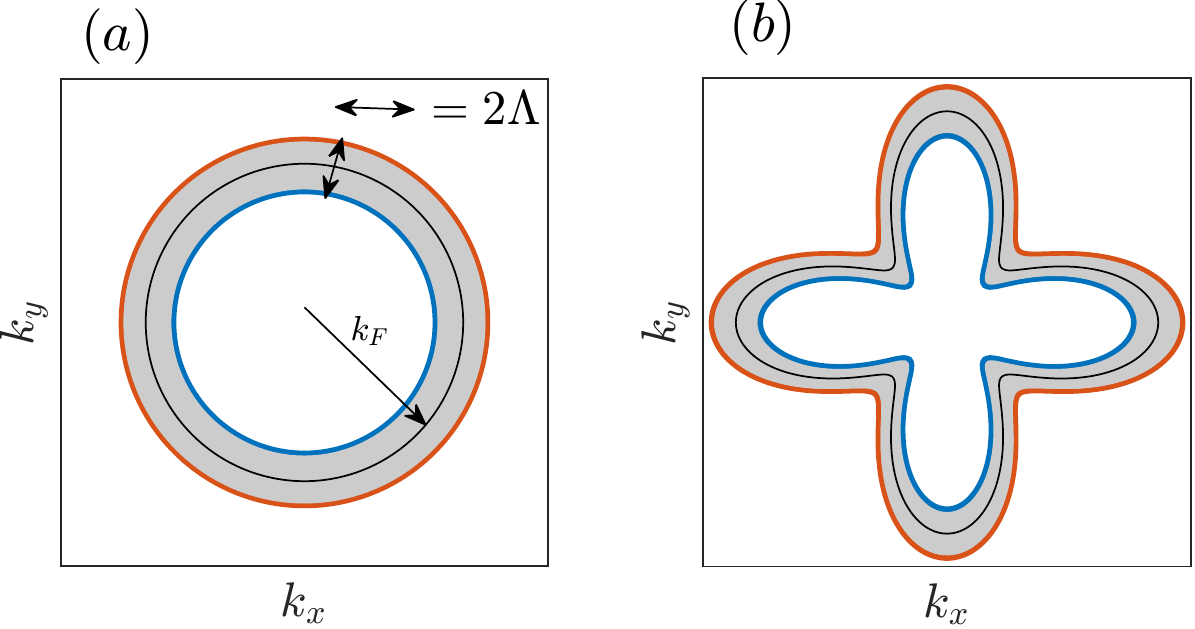}
	\caption{A schematic plot of the fast modes integration out for a time-reversal symmetric model. The shaded region around the Fermi surface (FS) represents the slow momenta while all the rest of the Brillouin zone (fast modes) is integrated out.  In \textbf{(a)} it is shown for the simple case of circular FS, \textbf{(b)} is for a generic case in which $\epsilon_{\mathbf{k}}=\epsilon_{-\mathbf{k}}$.   }
\label{fig:TR_fig}
\end{figure}

\underline{Step 1}: We divide the fermionic modes into slow modes residing within a thin shell of thickness $2\Lambda$ around the FS (shaded region in Fig. \ref{fig:TR_fig}), and fast modes residing outside of the shell. The fast modes are integrated out, obtaining the infra-red effective theory ($S_{\rm{IR}}$)  close to the FS.
 Note that $S_{\rm{IR}}$ is still quadratic in the fermions at this stage, as we do not integrate over $\phi$ at this stage~\cite{Note2}.

The infra-red action is written as
\begin{align}
\label{eq:SIR}
    S_{\rm{IR}}&=\int_{\left| \mathbf{\epsilon_{{k},\alpha}}\right| <\Lambda}   \mathrm{d}{k} \ \overline{\gamma}_{\alpha,{k}}
  \frac{\left(i\omega -h_{\alpha,\alpha',\mathbf{k}}  \right)}{2}  \gamma_{\alpha',{k}} 
+S'_{\phi}+S'_{\rm{Int}}
\end{align} 
with
\begin{align}
    S'_{\rm{Int}}=\int \mathrm{d}k \mathrm{d}k'\,\,
      \overline  \gamma_{\alpha,{k}} 
          \Sigma_{\alpha,\alpha',{k},{k'}}(\phi) {\gamma}_{\alpha',{k'}} \ .
\end{align}
Here, $\Sigma(\phi)$ is the fermionic self-energy obtained from integrating out the fast modes, including all $\lambda$ dependent terms (we suppress the indices of $\phi^{\nu}$ for brevity). $S'_{\phi}$ is the renormalized bosonic action. More details concerning the diagrammatic representation of $\Sigma(\phi)$ and $S'_{\phi}$ can be found in the SI~\cite{Note2}.
Importantly, $\mathbf{\epsilon}_{\mathbf{k}}$ is the ``true'' (renormalized) dispersion, given by the eigenvalues of the matrix $h_{\alpha,\alpha',\mathbf{k}}+\left<\Sigma_{\alpha,\alpha',{k},{k}}\right>_{S'_\phi}$ (with $\omega=0$ within the self-energy). Hence, the exact FS is given by $\mathbf{\epsilon}_{\mathbf{k}}=0$.

Next, we integrate out $\phi$ in order to obtain a purely fermionic low-energy action. By our assumption, this action has a FL form. We assume that we are not exactly at a quantum critical point (QCP);  
at a QCP, singular interactions between the low-energy fermions arise, violating the FL assumption.  

In the resulting fermionic action $S_{\rm{eff}}(\gamma,\overline{\gamma})$, we are interested only in the quartic terms, as in a FL, higher order terms are irrelevant in the renormalization group (RG) sense~\cite{Shankar1994,Polchinski1992}. 
The quartic part of $-S_{\rm{eff}}(\gamma,\overline{\gamma})$ is written as
\begin{equation}
    \Gamma^{\alpha,\beta,\alpha',\beta'}_{{k_1},{k_2},{{k}'_1},{{k}'_2}}
      \overline {\gamma}_{\alpha,{k}_1} 
       \overline {\gamma}_{\beta,{k}_2} 
     \gamma_{\beta',{k}'_2}  
     \gamma_{\alpha',{k}'_1}\delta(\{k\}), 
\end{equation}
where $\delta(\{k\}) = \delta_{k_1+k_2-k_1'-k_2'}$, and 
we can neglect the dependence of $\Gamma$ on both frequency and the momentum perpendicular to the FS, since both dependencies are irrelevant~\cite{Shankar1994,Polchinski1992}. 
Using the cumulant expansion, we can express $\Gamma$ as
\begin{align}
  \Gamma^{\alpha,\beta,\alpha',\beta'}_{{\mathbf{k}_1},{\mathbf{k}_2},{\mathbf{k}'_1},{\mathbf{k}'_2}}&= 
  \left<{\Sigma}_{\alpha,\alpha',{\mathbf{k}_1},{\mathbf{k}_1'}}{\Sigma}_{\beta,\beta',{\mathbf{k}_2},{\mathbf{k}_2'}} \right>_{S'_{\phi}} \\ \notag
  &-  \left<{\Sigma}_{\alpha,\alpha',{\mathbf{k}_1},{\mathbf{k}_1'}} \right>_{S'_{\phi}} \left<{\Sigma}_{\beta,\beta',{\mathbf{k}_2},{\mathbf{k}_2'}} \right>_{S'_{\phi}}.
  \label{eq:Gamma}
\end{align}
 Note that, due to the SF property of the model, $S'_{\phi}\in \mathbb{R}$(see~\cite{Note2}).
 
 \underline{Step 2}:
 We now perform a stability analysis of the putative FL fixed point. 
 The analysis follows closely the standard RG procedure for interacting fermions~\cite{Shankar1994}. In the limit $\Lambda\rightarrow 0$, forward scattering processes are exactly marginal. Since the model is SF, it has at least one TRS, and hence $\epsilon_{\mathbf{k}} = \epsilon_{-\mathbf{k}}$. 
 Then, FS instabilities may arise in the Cooper channel, corresponding to
 ${k}_1=-{k}_2, {k}'_1=-{k}'_2$. 
 To obtain a compact form of the RG equations, it is convenient to define  
\begin{equation}
\tilde{\Gamma}^{\alpha,\beta,\alpha',\beta'}_{\mathbf{k},-\mathbf{k},\mathbf{k}',-\mathbf{k}'} \equiv \frac{1}{\sqrt{v_\mathbf{k}  v_{\mathbf{k}'}}} \Gamma^{\alpha,\beta,\alpha',\beta'}_{\mathbf{k},-\mathbf{k},\mathbf{k}',-\mathbf{k}'} \ 
\end{equation}
where $v_\mathbf{k} = |\nabla_{\mathbf{k}} \epsilon_{\mathbf{k}}|$. 
We treat $\tilde{\Gamma}^{\alpha,\beta,\alpha',\beta'}_{\mathbf{k},-\mathbf{k},\mathbf{k}',-\mathbf{k}'}$ as a matrix $\tilde{\mathbf{\Gamma}}$,  where the first (second) index corresponds to the set $\{\alpha,\beta,k\}$ ($\{\alpha',\beta',k'\}$), respectively. The hermiticity of the effective FL Hamiltonian implies that the matrix $\tilde{\mathbf{\Gamma}}$ is hermitian. 

The one-loop RG equations for $\tilde{\mathbf{\Gamma}}$ take the simple form 
\begin{equation}
\frac{d\tilde{\mathbf{\Gamma}}}{dl} = \frac{1}{4\pi}\tilde{\mathbf{\Gamma}}^2,   
\end{equation}
where $dl=-d\Lambda/\Lambda$ is the infinitesimal scaling factor. 
Diagonalizing $\tilde{\mathbf{\Gamma}}$, we obtain a set of differential equations for the eigenvalues (denoted by $\lambda_{i}$): 
\begin{equation}
\frac{d\lambda_i}{dl} =\frac{1}{4\pi} \lambda_{i}^2 .
\label{Eq:lambda_bcs}
\end{equation}
A positive $\lambda_i$ (corresponding to attraction in a certain channel) grows under RG, destroying the FL. Thus, a stable FL phase requires that $\lambda_i<0$ for all $i$. Conversely, if there is at least one positive eigenvalue, there is no FS at $T=0$. If $\lambda_i\le 0$ for all $i$, but there is at least one zero eigenvalue, there can be a FS, but it is unstable to the addition of an infinitesimal perturbation that makes one of the eigenvalues positive. (The latter case corresponds, e.g., to a non-interacting electron gas). 

We now show that there exists a vector $\vec{w}$ for which $\vec{w}^T \tilde{\mathbf{\Gamma}} \vec{w} \ge 0$, and hence $\tilde{\mathbf{\Gamma}}$ has at least one non-negative eigenvalue, and the FL phase cannot be stable.  
As mentioned above, the model has at least one TRS.  
We write the corresponding TRS operator as $T=OK$, where $O$ is an orthogonal matrix ($O$ is real since we are dealing with Majorana fermions), and $K$ denotes complex conjugation. Under TRS, $\gamma_{\alpha,{k}} \to O_{\alpha\beta}\gamma_{\beta,{-k}}$. Setting $w_{\mathbf{k},\alpha,\beta}=O_{\alpha\beta} \delta_{\mathbf{k},\mathbf{k}_0}$ with an arbitrarily chosen $\mathbf{k}_0$, and using the identity 
\begin{equation}
\sum_{\alpha,\alpha'} O_{\alpha\beta}\Sigma_{\mathbf{k}_{0},\mathbf{k}_{0}}^{\alpha,\alpha'}O_{\alpha'\beta'}=\left(\Sigma_{-\mathbf{k}_{0},-\mathbf{k}_{0}}^{\beta,\beta'}\right)^{*},
\end{equation}
which follows from the time-reversal invariance of the fermionic action under $T$, we obtain
\begin{align}
\vec{w}^{T}\tilde{\mathbf{\Gamma}}\vec{w}	&=\frac{1}{v_{\mathbf{k_{0}}}}\sum_{\beta,\beta'}\Big[\left<\left|\Sigma_{\beta,\beta',-{\mathbf{k}_0},-{\mathbf{k}_0}}\right|^{2}\right>_{S'_{\phi}} \notag\\
	&-\left<\Sigma_{\beta,\beta',-{\mathbf{k}_0},-{\mathbf{k}_0}}\right>_{S'_{\phi}}\left<\Sigma_{\beta,\beta',-{\mathbf{k}_0},-{\mathbf{k}_0}}^{*}\right>_{S'_{\phi}}\Big]\geq 0.
 \label{ineq}
 \end{align}
 Thus, the putative FL phase is either intrinsically unstable, or can be destabilized by adding an infinitesimal attractive interaction. This is our main result.

\begin{figure}
        	\centering\includegraphics[ width=1\columnwidth ]{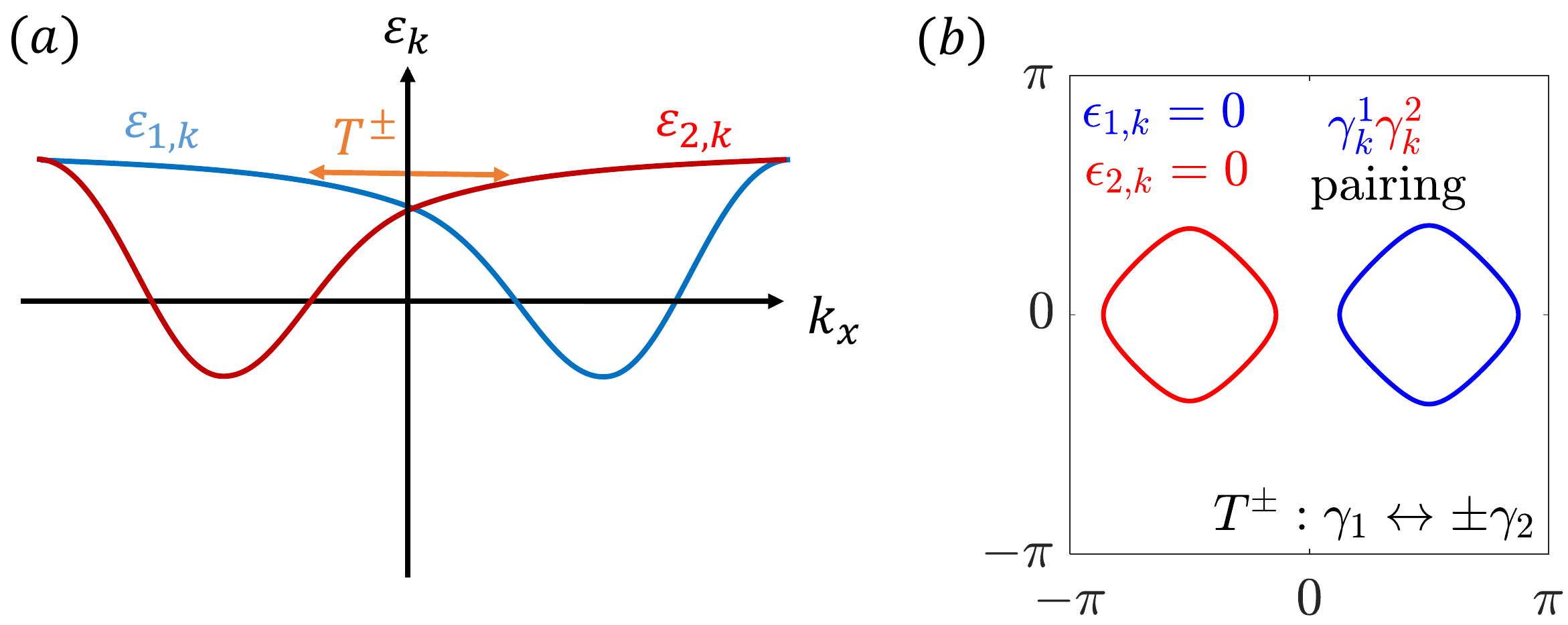}
    	\caption{Schematic description of Cooper channel instability in the sign-free  classes. We demonstrate it for the simple case of 2 flavors. (\textbf{a}) The time-reversal mapping between the two bands.  (\textbf{b}) The obtained nesting between the Fermi surfaces. }
\label{fig:TRS_nesting}
\end{figure}

It is worth examining the key elements required for our proof. In essence, the TRS (which all presently known SF classes require) ensures that the FS has a Cooper-like instability, with states at opposite momenta being degenerate 
(see Fig.~\ref{fig:TRS_nesting}).  
In addition, the SF property guarantees that the effective interaction is attractive in some channel (although it may be repulsive in other channels). Therefore, the FS cannot be stable. It is natural to expect a gapped, spontaneously broken ground state as a result.  

We further stress that we cannot rule out the possibility of a non-FL metal with a FS. 
We also do not rule out a manifold of co-dimension $d-2$ or lower of zero energy excitations. SF models can have a stable Dirac semi-metal ground state in $d=2$~\cite{meng2010quantum,sorella2012absence,Gazit2017,Tang2018}, or a nodal line semimetal in $d\geq3$.

Importantly, within our proof, we have implicitly assumed that TRS is not broken \emph{spontaneously}~\footnote{Importantly, here we refer to breaking of all anti-unitary symmetries. In the symmetric SF classes, it is possible that one of the protecting TRSs is spontaneously broken, while the other is preserved.}. If such spontaneous symmetry breaking had occurred, the FS could be stable. Ref.~\cite{Ringel2017} conjectured, and showed explicitly for some cases, that an anti-unitary symmetry cannot be spontaneously broken in a bosonic SF model. We do not know of cases where TRS is spontaneously broken in a SF model containing fermions; whether such TRS breaking is fundamentally possible remains to be seen. 

{\it The gapped phase.---}
While for all SF classes, the ground state is not a stable FL, its exact nature is class and model-dependent. 
However, in both symmetry classes, it is natural to expect that as a result of the FS instability, the unitary $T_1 T_2$ symmetry is spontaneously broken, and the FS is gapped out. In Kramers' class, this typically results in a superconducting ground state, as $\gamma^{T} iT_1 T_2\gamma$ is the generator of a $U(1)$ symmetry. The angular momentum of the order parameter is model specific, and both $s-$wave~\cite{Scalettar1989,Moreo1991,Noack1991,Schattner2016,Loh2016,Esterlis2018,Hofmann2020} and nodal or nodeless $d-$wave~\cite{Assaad1996,Schattner2016a,Xu2021} superconductivity were found within SF DQMC.    
 In the Majorana class, $T_1T_2$ is a unitary  $\mathbb{Z}_2$ symmetry. Its spontaneous breaking results in a 2-fold degenerate ground state.  
 In certain physical realizations, the symmetry broken phase may correspond to a charge ordered state~\cite{Noack1991,Johnston2013,Li2015,Wei2016,Kivelson2017,Esterlis2018}. 

It is important to note that the spontaneous breaking of $T_{1}T_{2}$ does not always occur; if the fermions form a band insulator or a Dirac semi-metal, the symmetry unbroken phase may extend down to $T=0$. We provide detailed examples of realizations in different symmetry SF classes and their symmetry broken phases in ~\cite{Note2}.
 
 In the lower-symmetry SF classes, as we only have a single TRS and not a unitary symmetry, 
 there is generically no FS already at the level of the quadratic (non-interacting) part of the action. 
 The single-particle Hamiltonian is analogous to a Bogoliubov-de Gennes (BdG) Hamiltonian with TRS, where real off-diagonal pairing terms are allowed. Consequently, the co-dimension of the zero energy modes is at most $d-2$ 
 (see~\cite{Note2}). If we add a symmetry (beyond those required by the SF property) that protects the FS at the quadratic level (e.g., a $U(1)$ or $\mathrm{Z}_2$ symmetry), the FS is still generically unstable in the presence of interactions, just as in the symmetry classes. This is because our proof requires only a single TRS, which is present in the lower-symmetry classes.   
 We summarize our conclusions for all currently known SF classes in Table \ref{table:classes}.

 {\it Comment regarding the Kohn-Luttinger mechanism.--}
We have shown that SF models cannot have a stable FL ground state, due to TRS and the presence of attraction in the Cooper channel. In this context, it is important to address the question whether, conversely, non-SF models with TRS can support a stable Fermi surface at $T=0$. It is well known that under a wide range of circumstances, bare repulsive interactions can lead to superconductivity at high angular momentum channel. This is the Kohn-Luttinger (KL) mechanism~\cite{Kohn1965}.

In general, however, the KL mechanism relies on the bare repulsion being short-ranged, i.e., exponentially decaying as a function of distance (see~\cite{Note2} for details). The Fermi surface can be stabilized at $T=0$ by adding a small power-law repulsive interaction. In contrast, our proof shows that in SF models, the Fermi surface is generically unstable, even in the presence of arbitrary long-ranged interactions. 

 {\it Concluding remarks.--} 
 Our arguments apply to all  SF classes that are currently known within DQMC.  However, these observations naturally lead to the stronger conjecture that a stable FL phase cannot be realized in any SF model. As new classes of SF models are found, this conjecture will be put to the test. For example, there are models~\cite{Huffman2016} that do not have a known SF DQMC formulation, but the sign problem can be solved in a continuous time QMC~\cite{Gull2011,Wang2014,Wang2015,Huffman2014} (or the ``fermion bag'' approach~\cite{Chandrasekharan2010,chandrasekharan2013}). None of these models have a FL ground state, as far as we know; however, our proof does not formally encompass these cases. In addition, we note that it is possible to get stable metallic phases in ``mixed dimensioanlity" systems~\cite{Frank2022,Frank2022B}, where the interaction terms are sub-extensive. These lie beyond the scope of our results since they are not fully translationally invariant.

Looking ahead, it should be straightforward to extend our results to include quenched disorder. In this case, we expect SF models to obey a version of Anderson's theorem~\cite{Anderson1959}, i.e., disorder that preserves the SF property does not suppress the superconducting instability at the mean-field level. 

 {\it Acknowledgments--}
We are grateful to A. Chubukov, R. Shankar, S. Gazit, O. Golan, S. Kivelson, D. Mross, Z. Ringel, J. S. Hofmann and S. Trebst 
for illuminating discussions. 
This work was supported by the European Research Council (ERC) under the European Union’s Horizon 2020 research and innovation programme (grant agreement No 817799), the Israel-USA Binational Science Foundation (BSF), and the ISF Quantum Science and Technology grant no. 2074/19. 

\bibliography{main.bbl}

\clearpage

\appendix
\onecolumngrid
\textbf{\center\large{Supplementary Information: Robust Fermi liquid instabilities in sign problem-free models}}
\\
\\
In this Supplementary Information, we provide a detailed explanation regarding some of the proof steps. In addition, we provide examples for the different sign problem-free symmetry 
 classes and elucidate more about the lower-symmetry SF classes. Lastly, we discuss the value of our approach in the context of Kohn-Luttinger mechanism.

\section{Examples for the symmetry sign-free classes} 
\label{App:examples}
In this section, we briefly mention  some familiar examples of SF models and their FL instability. We start with the canonical example of Hubbard model as a representative model of Kramers' class. Consider the Hamiltonian 
\begin{equation}
\hat{H}_1=-\sum_{i,j}\sum_{\sigma}(t_{ij}c^{\dagger}_{i,\sigma}c_{j,\sigma} +\textnormal{H.c})+U\sum_{i\in\Lambda}c^{\dagger}_{i,\uparrow}c^{\dagger}_{i,\downarrow}c_{i,\downarrow}c_{i,\uparrow}   .
\label{eq:hubbard}
\end{equation}
The Hubbard model is known to be SF for attractive interaction ($U<0$) at any chemical potential $\mu$, and  for repulsive interaction ($U>0$) if the lattice is bipartite and $\mu=0$. Performing HS transformation in the density channel for the attractive case (see ~\cite{wu2005}) and using Majorna representation ($c=\frac{\gamma_1 + i\gamma_2}{2}$), we can easily find that $T^{-}_1=i\sigma_y\tau_z K$ and $T^{-}_2=i\sigma_y \tau_x K$ are mutually anti-commuting TRSs. Here, $\sigma_{x,y,z}$ act on the spin degree of freedom and $\tau_{x,y,z}$ acts in the Majorana sector ($\gamma_{1,2}$). The continuous $U(1)$ global symmetry generated by $T_1T_2=\tau_y$  is nothing but the conservation of charge.
As expected, the SF property (for $\mu\ne 0$) eventually leads to an emergent SC gap and breaking of the global $U(1)$ symmetry upon cooling. 

In the repulsive case, decoupling again in the density channel, we find that   $T^{-}_1=i\sigma_y \eta_{i} K$ and $T^{-}_2=i\sigma_x \tau_y \eta_{i} K$ with $\eta_{i}=(-1)^i$ such that $i$ is even (odd) for sublattice $A$ (sublattice $B$). 
In this case, the continuous global symmetry $T^{-}_1 T^{-}_2=\sigma_z\tau_y$ (which is the generator of spin rotations around $\hat{z}$) can be spontaneously broken in a Mott insulating phase with antiferromagnetic order. Since the model has $SU(2)$ symmetry, ground states that preserve this symmetry also exist (recall that our proof does not imply that $T_1T_2$ must always be broken - but only that the FS is always unstable).

We find a similar behavior also in the Majorana class. Let us consider the spinless $t-V$ model:
\begin{equation}
\hat{H}_2=-t\sum_{\left<i,j\right>}(c^{\dagger}_{i}c_{j} +\textnormal{H.c})+V\sum_{\left<ij\right>}(\hat{n}_i -\frac{1}{2})(\hat{n}_j - \frac{1}{2})
\end{equation}
  on a bipartite lattice and with $V>0$. Using the Majorana representation, we find that the TRSs are $T_{1}^{-}=i\tau_y \eta_{i} K$ and $T_2^{+}=\tau_x\eta_{i}K$, where $\tau$ acts on the Majorna sector and $\eta_{i}$ is defined as before. Note 
  that these symmetries are mutually anti-commuting and that their product $T_1T_2=\tau_z$ takes $\gamma_1\rightarrow \gamma_1$ and $\gamma_2\rightarrow -\gamma_2$. Physically, $T_1T_2$ interchanges particles and holes. This particle-hole symmetry is spontaneously 
  broken into a CDW phase that emerges at large $V/t$ ~\cite{Li2015}.

\section{Technical details of the proof} 

Here we elaborate on various steps of the proof that require further explanation.
\label{appendix:proof}
\subsection{Integrating fast fermionic degrees of freedom }
\label{appendix:proof_1}
\begin{figure}[h]

        	\centering\includegraphics[ width=0.5\columnwidth ]{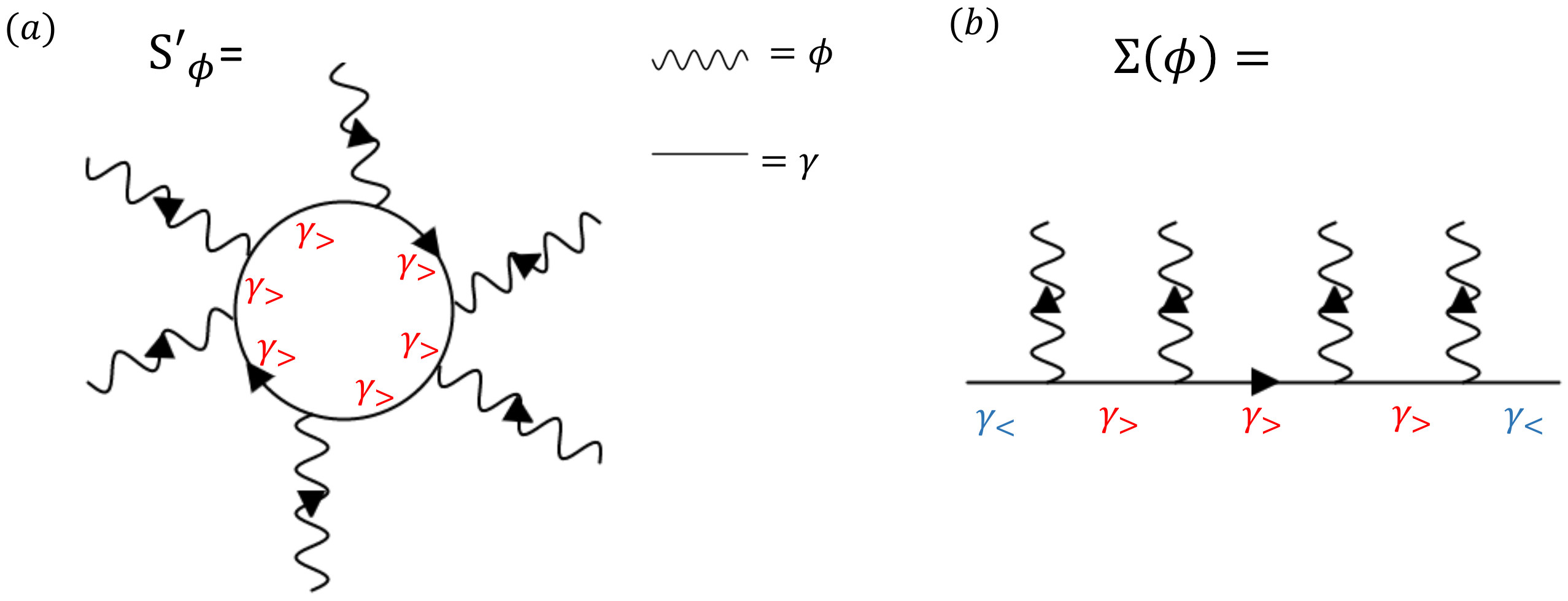}

	\caption{An exemplar  diagram for each kind of the terms that we generate when tracing out the fast modes in our $UV$ model.  (\textbf{a}) The type of diagrams that  contribute to the effective bosonic action $S'_{\phi}$ (\textbf{b}) The correction for the self-energy  $\Sigma(\phi)$. We cannot generate higher-order fermionic terms, since we do not trace over bosons (no internal bosonic lines).  }
\label{fig:psi_fast_diagrams}
\end{figure}

When integrating out the fast fermionic degrees of freedom $\gamma_{>}$ (whose momenta are outside the thin shell around the FS), we need to consider two kinds of terms that are generated, corresponding to the diagrams shown in Fig. \ref{fig:psi_fast_diagrams}: 
\begin{enumerate}[label=(\alph*)]
  \item The loop diagrams, which contribute to the effective bosonic action
$S'_{\phi}$.
  \item The ``comb'' diagrams, which contribute to the self energy correction $\Sigma(\phi)$.  
\end{enumerate}
 Note that $\Sigma(\phi)$ has only ``comb'' diagrams, since we do not integrate over the bosonic degrees of freedom. For the same reason, no quartic fermionic terms are generated at this step.

\subsection{The Sign-free property in the low energy model}
\label{appendix:proof_2}
Let us consider the Majorana fermionic action matrix $\tilde{M}_{\phi}$ in momentum space:
  \begin{align}
      \tilde{M}_{\phi}= \begin{bmatrix}
        M_< &  M_{i} \\  -M^T_{i}    & M_> 
      \end{bmatrix}, 
  \end{align}
where the blocks of $ M_< $ and  $M_>$ act on the slow and fast degrees of freedom, respectively (in general, these blocks do not have the same size), and $ M_{i}$ (which originates from the interaction with the bosons) couples off-diagonal blocks. Note that in the off-diagonal blocks, we have used the skew-symmetric property of $\tilde{M}_{\phi}$. When completing to square, We find that 
  \begin{equation}
{\gamma}_{<}M_{<}\gamma_{<} +
{\gamma}_{>}M_{>}\gamma_{>}+
2{\gamma}_{<}M_{i}\gamma_{>}
=
({\gamma}_{>}+{\gamma}_{<}M_{i}M^{-1}_{>})M_{>}({\gamma}_{>}-M^{-1}_{>}M^{T}_{i}{\gamma}_{<}) +{\gamma}_{<}M_{i}M^{-1}_{>}M^T_{i}\gamma_{<}+{\gamma}_{<}M_{<}\gamma_{<},
  \end{equation}
  where, for simplicity, we do not display the momentum labels. 
  After tracing out the fast modes,  the effective bosonic action is given by $S'_{\phi}=S_{\phi}+\ln\det M_{>}$, and the remaining fermionic action is $S'_{\gamma}={\gamma}_{>}M_{\rm{eff}}{\gamma}_{>}$, where $M_{\rm{eff}} \equiv M_{<}+ M_{i}M^{-1}_{>}M^T_{i}$. We now show that $M_{>}$ belongs to the same SF class as $\tilde{M}_{\phi}$ (hence $\det M_{>}\geq 0$ and $S'_{\phi}\in\mathbb{R}$).

For all SF classes, there is one TRS  ($T_{2}^{-}$ for Kramers classes, $T_{2}^{+}$ for Majorana classes) which satisfies $\left[ T_{2}^{\pm} ,\tilde{M}({\phi})\right]=0$, and a second TRS, $T_{1}^{-}$ ( where $\left\{T_{1}^{-}, T_{2}^{+} \right\}=0$), which satisfies $iK\left[ T_{1}^{-} ,\tilde{M}({\phi})\right]  \leq 0$. 
Denoting $T_{1}^{-}=OK$, where $O$ is the orthogonal matrix $O=\textrm{diag}(O_<,O_>)$, we find that
  \begin{align}
  iK\left[ T^{-},\tilde{M}({\phi})\right] =      i\begin{bmatrix}
        O_<M_<  -  M^*_<O_< & O_<M_i - M^*_iO_> \\    M^\dagger_iO_< -O_>M^T_i  &   O_>M_>  -  M^*_>O_> 
     \label{Eq:UV-NS}
      \end{bmatrix} \leq 0 \ .
  \end{align}
Note that the above inequality holds in particular for each of the diagonal blocks (this can easily be shown by using the negative semi-definite property for an arbitrary vector whose non-zero entries are only in the fast or slow modes).
Hence,  $iK\left[ O_{>}K ,\tilde{M}_>\right]  \leq 0$ and therefore $M_{>}$ is SF. In addition, since 
$\det\tilde{M}_{\phi} = \det M_{>} \det M_{\rm{eff}} $, it follows that $\det M_{\rm{eff}} \geq 0$ and the remaining action is also SF, as expected. It is also important to note that $T^{\pm}_{2}$ is a symmetry of the effective action, as it does not mix slow and fast momenta.

\section{Absence of Fermi surface in the lower-symmetry classes}
\label{Appendix:NS_class}
In this section, we briefly explain why in the lower-symmetry SF classes, there is generically no Fermi surface (unless additional symmetries, not related to the ones necessary for the SF property, are imposed). Intuitively, the low-energy quadratic effective Hamiltonian resembles a Bogoliubov-de Gennes Hamiltonian of a mean-field superconductor, in which pairing terms are allowed. Time reversal symmetry guarantees that the gap function can be chosen to be real. In this case, the Fermi surface is generically gapped, with the exception of possible nodal points in two dimensions (or nodal lines in three dimensions). 

To show this, we start by considering the Majorana lower-symmetry class. In this case there is only one TRS, $T_{2}^{+}$, such that  $\left[T_{2}^{+},\Tilde{M}_{\phi}\right]=0$. An additional anti-unitary operator $T_{1}^{-}$ anti-commutes with $T_2^{+}$ and fulfils the inequality condition $iK\left[ T_{1}^{-} ,\tilde{M}_\phi \right]  \leq 0$. We can construct a basis for $\Tilde{M}_{\phi}$ labelled by the eigenvalues $\pm 1$ of the parity operator, $T_{1}^{+}T_{2}^{-}$. 
We choose $S\equiv T^{+}T^{-}=\sigma_{z}$.  
In this basis, the two TRS operators can be represented as $T_1^{-}=i\sigma_y K, T_2^{+}=\sigma_x K$.

We now assume that there is a Fermi surface and study its stability. Following the steps leading to Eq.~(\ref{eq:SIR}) in the main text, we consider the quadratic part of the low-energy effective Hamiltonian, $h_{\mathbf{k}}$. This effective Hamiltonian must anti-commute with $T_{2}^+ K = \sigma_x$. Hence, each zero-energy eigenstate must be at least two-fold degenerate. Near such a degeneracy point, $h_{\mathbf{k}} \approx a_{\mathbf{k}} \sigma_y + b_{\mathbf{k}} \sigma_z$, where $a_{\mathbf{k}}$ and $b_{\mathbf{k}}$ are real functions. Generically, the co-dimension of the zero energy manifold is $d-2$, since a degeneracy point requires $a_{\mathbf{k}} = b_{\mathbf{k}} = 0$. Hence, the Fermi surface is generically unstable. Only Dirac point nodes in $d=2$ (or line nodes in $d=3$) are generically allowed. 

The argument is similar for the Krammers' lower-symmetry class. In this case, it can be shown that pairing terms are allowed when there is only one TRS, $T_{2}^{-}$ (see ~\cite{Carsten_thesis_phd} for more details) and we also get a Bogoliubov de-Gennes Hamiltonian which is typically gapped.

\section{Beyond Kohn-Luttinger superconductivity }
 \label{appendix:KL_machanism}
 The well-known work of Kohn and Luttinger (KL)~\cite{Kohn1965} showed that even systems with bare repulsive interactions can become superconducting at sufficiently low temperatures.
Here, we recapitulate the
KL mechanism, using a simple model as an illustration. 
We show that the KL mechanism is not necessarily applicable when long-range interactions are present, and in general, a stable Fermi surface may exist, even when time reversal symmetry is present. In contrast, SF models cannot have stable Fermi surfaces either for short or long-range interactions. 

 We consider the following model in two spatial dimensions:  
 \begin{equation}
\hat{H}=\sum_{\mathbf{k},\sigma}
\epsilon_{\mathbf{k}} c^{\dagger}_{\mathbf{k},\sigma}c_{\mathbf{k},\sigma}
+ \sum_{\mathbf{k},\mathbf{k}',\mathbf{q}}
U_{\mathbf{k},\mathbf{k}',\mathbf{q}}
c^{\dagger}_{\mathbf{k'-q},\uparrow}
c^{\dagger}_{\mathbf{k+q},\downarrow}
c_{\mathbf{k},\downarrow}
c_{\mathbf{k}',\uparrow}
 \end{equation}
 where the interaction is assumed to be short-ranged in real space, and therefore $U_{\mathbf{k},\mathbf{k}',\mathbf{q}}$ is a smooth function of its arguments. For simplicity, we assume that the problem is rotationally symmetric. We correspond the interaction in the Cooper channel, setting $\mathbf{k}=-\mathbf{k'}$. Because of rotational symmetry,  $U_{\mathbf{k},-\mathbf{k},\mathbf{q}}$ depends only on the momentum transfer $\mathbf{q}$ in the low energy limit, where we set $|\mathbf{k}|=|\mathbf{k'}|=k_F$. We can therefore parameterize $U_{\mathbf{q}}$ by the relative angle $\theta$ between $\mathbf{k}$ and  $\mathbf{k+q}$, such that $|\mathbf{q}|=2k_F \sin\frac{\theta}{2}$.
  The interaction is diagonal in the angular momentum basis. We denote the $m$th Fourier component of the interaction by $\Tilde{U}_m$.  Up to second order in the original interaction, the two-particle vertex in the Cooper channel is given by
\begin{align}
    \Tilde{U}_{\rm{eff},m} =\tilde{U}_m-\frac{|C|\nu_0 \Tilde{U}_{\pi}^2}{m^2}.
        \label{Eq:bcs}
\end{align}
Here, $\nu_0$ is the density of states at the Fermi level, and $C$ a number of order unity. The first term in the right-hand-side, $\tilde{U}_m$, decays exponentially with $m$ if the original interaction is repulsive and short-ranged. 
On the other hand, the second term is negative (attractive) in sign, and decays as $1/m^2$. This behavior originates from the Linhard function of the Fermi gas at $\theta=\pi$ ( i.e, when $q=2k_F \sin\frac{\theta}{2}=2k_F$). A detailed 
derivation of this result appears in Ref.~\cite{Chubokov1993}. In particular, the coefficient $C$ vanishes if the dispersion $\epsilon_{\mathbf{k}}$ is parabolic, but is generically non-zero for a general dispersion. It is thus clear that for large enough $m$, the power law term is dominant over the exponentially small term, and the net interaction is attractive, leading to an instability towards superconductivity. This is the idea behind the KL mechanism.

However, the above argument may fail if the bare interaction is long-ranged. Consider, for example, a Thomas-Fermi screened Coulomb potential in $d=2$. In this case, the potential takes the form of $U(q)=\frac{2\pi e^2}{|\mathbf{q}|+2\pi e^2 \nu_0  }$, where $e$ is the electron charge. The singularity of $U(q)$ at $q=0$ leads to a power law decay $\tilde{U}_m \sim m^{-2}$ at large $m$. This is the same power law as that of the second term in Eq.~\eqref{Eq:bcs}; hence, in this case, there is no reason to assume that the attractive becomes dominant, even at arbitrarily large $m$. Therefore, in this case, the KL mechanism does not necessarily lead to superconductivity. In contrast, in a SF model, the effective interaction is always attractive in a certain channel, irrespective of whether it is short or long ranged in space. 

We note that the renormalization group (RG) treatment is not applicable 
in the case of long-range bare interaction. 
In addition, Eq.~\eqref{Eq:bcs} cannot be rigorously derived from RG (since the $1/m^2$ dependence of the second term comes from fermions that are arbitrarily close to the Fermi surface, and hence cannot be obtained by integrating out fermions away from the Fermi surface). 
Nevertheless, we believe that our argument that adding a weak long-range interaction can prevent the KL instability and stabilize the Fermi surface in a non-SF model is highly suggestive (though not strictly rigorous).

\end{document}